\documentstyle[12pt]{article}\textwidth 17cm
\begin{document}
\large

\vspace{40mm}
\begin {center}
{\LARGE
DETECTORS FOR THE COSMIC\\
AXIONIC WIND
}
\end {center}
\vspace{0mm}
\begin {center}
{\em P.V.Vorobyov$^*$,
I.V.Kolokolov$^*,^{**}$}\\$ ^*$Budker Institute of Nuclear
Physics,\\ Novosibirsk 630090, Russia
\\$^{**}$INFN, Sez.di Milano, 20133 Milano, Italia
\end {center}
\vspace{0mm}
\begin{quotation}{\em
We propose experimental schemes for detection an axionic
condensat supposed to be a cosmic dark matter. Various procedures
are considered in dependence on the value of the axion mass.
}
\end{quotation}
\vspace{4mm}

\section{Introduction.}
There are well known indications that a large part of the Universe
mass exists in a form of dark matter:

The analysis of rotation curves of galaxies leads to the conclusion that the
mass of luminous matter is less than 1/10 part of the total galaxies mass
\cite{Vor},\cite{Ei}.

The existence of the dark matter is supported by the so called "virial
paradoxes". In a stationary system the virial theorem gives the relation:
\begin{equation}\label{virial}
E+2P=0,\;\; or \;\;M\overline{\bf v^2}+P=0,
\end{equation}
where $E$ is the kinetic energy of the system, $P$ is the potential
energy, $M$ is its mass and $\overline{\bf v^2}$
is the velocity variation.
It turns out that the reach and compact galaxies have inacceptible
large $\overline{\bf v^2}$  being in the same time stable with respect
to anothers characteristics. For such the galaxies to be stable their
masses must be one order greater than the observable ones
\cite{Vor},\cite{G}. There
are theoretical and observational arguments that this
dark matter cannot be usual barionic matter as dust, planets etc.

On the other hand there are attractive models where the dark matter
is non-relativistic gas of light elementary particles weakly interacting
with the "usual" matter \cite{Ell}, \cite{Khl1}.
Periodicity in the distribution
of quasars and distant galaxies with the red shift \cite{Petja1},
\cite{Bro}
could be naturally
explained in the cosmology with a gas of very light (pseudo)Goldstone
bosons filling the Universe \cite{Ans},\cite{M}.
Small mass and small interaction constants are ordinary properties
of pseudoscalar particles arising in various Grand Unification schemes
(axions, familons, arions etc.).

In the present paper we describe some ways to detect the presence of
such a pseudoscalar surroundings \cite{VKK}.
We will speak the corresponding
particles as axions not assuming, however, a strict relation between
the mass and the coupling constant.

\section{The axionic wind.}
It is well known that during a cooling an ideal Bose gas undergo the
Bose condensation. The value of the transition temperature in Kelvin
can be found as
\begin{equation}\label{condens}
T_0=\frac{5\cdot 10^{-3}}{m^{5/3}} \left(\frac{\rho}{\rho_c}\right)
^{2/3}.
\end{equation}
Here $m$ is the mass of a gas particle expressed in $eV$, $\rho$ is the
density of the gas and $\rho_c$ is the critical cosmological density
being set here $10^{-29}g/cm^3$. If $m\leq 0.1 eV$ the value of $T_0$
is of the order of the relict radiation temperature. In the standard
cosmological model when the cosmological scale factor $R$ ("the radius
of the Universe") decreases, the temperature $T$ of the matter behaves
like $R^{-1}$ whereas $T_0\propto R^{-2}$ \cite{W}.
Thus, if the decoupling of axions has been enough early in the cosmological
evolution  the axion
gas should be in that moment of time
in the Bose condensat state. An adiabatic expansion cannot
destroy the ground state. Consequently, in  subsequent time the axions
would remain in the Bose condensat. The latter is described in the proper
frame of references by the macroscopic wave function:
\begin{equation}\label{w-func}
\Psi_0=\sqrt{n_a} e^{i\alpha}
\end{equation}
 Here $n_a$ is the number of axions per unit valume,$\alpha$ is some
constant phase.
The analysis given above used essentially the idealness of the axion gas.
However, pseudoscalar particles in various models can decay into two
photons, they  gravitate and can be emitted and absorbed by the matter.
It means that the axion condensat is in non-equlibrium state from the
very beginning. For the coupling constants of interest the thermalization
time exeeds the life time of the Universe. We can thus neglect the
influence of local interactions on the condensat. The gravitation can
lead to the adiabatic development of large-scales instabilities forming
space-inhomogenious distribution of axion density. In the present
paper we will not study the dynamics of axion condensat. We suppose
simply its presence in the vicinity of the Earth with some density.
We restrict here the discussion to the ways of detecting such an axionic
wind for various values of the axion mass.

The interaction of the axion field $\phi$ with the fermion one $\psi$
is described by the Lagrangean density:
\begin{equation}\label{L-int}
{\cal L}_{int}=iq_a \phi \bar {\psi}\gamma^5\psi
\end{equation}
$\psi$ is cosidered below as the electron field and $q_a$ is the
 dimensionless electron-axion coupling constant.

For the condensat state the expectation value of the field operator
$\phi(x)$ differs from zero.
(To avoid misunderstanings we stress that we consider the condensat
of $real$ particles which differs drastically from Lorentz invariant
condensats of the quantum field theory.)
In a frame of reference moving with a
non-relativistic velocity $\bf v$ through the condensat the value of
$<\phi>$ is equal to:
\begin{equation}\label{phi-aver}
<\phi>=\frac{1}{\sqrt{m_a}}\left(\Psi_0 e^{-im_at+im_a\bf v \bf x} +
\Psi_0^* e^{im_at-im_a\bf v \bf x}\right),
\end{equation}
The
electron-condensat interaction resulting from (\ref{L-int}) is taken into
account by adding to the electron Hamiltonian the term:
\begin{equation}\label{V-int}
\hat V=\mu_a\nabla <\phi>\mbox{\boldmath$\sigma$}=
i\mu_a\sqrt{m} ({\bf v} \mbox{\boldmath$\sigma$})
\left( \Psi_0 e^{-im_at
+im_a{\bf v}{\bf x}}-
\Psi_0^* e^{im_at
-im_a{\bf v}{\bf x}}\right),
\end{equation}
where $\mu_a=q_a/2m_e$ is the axionic magneton of the electron,
$m_a$ is the axion mass and $h/2\pi=c=1$ is assumed.
It means that a non-relativistic electron perceives the axionic
condensat as a spase-inhomogenious magnetic field oscillating in time.
The effective strength of this field is equal to:
\begin{equation}\label{Field}
{\bf B}_{eff}=2\kappa\sqrt{\rho_a}{\bf v}\sin
(m_at
+m_a{\bf v}{\bf x}+\theta).
\end{equation}
Here $\rho_a$ is the density of the condensat, $\kappa=\mu_a/\mu_B$,
$\mu_B$ is the Bohr magneton and $\theta$ is some phase.
Such an unusual manner of the matter---condensat interaction
results from the $field\times current$ form of
${\cal L}_{int}$ (\ref{L-int}) where just the field factor refers
to the condensat. (For example, the helium condensat is represented
in all interaction Lagrangians by  the current factor. A more appropriate
analogy is the scattering of a classical electromagnetic wave on an
electron.)

Let us suppose that ${\bf v}$ is equal to the "cosmological"
velocity of the Earth: $v\approx 10^{-3}$. Then the wavelength
corresponding to the space variations of the field ${\bf B}_{eff}$
can be estimated as
\begin{equation}\label{Length}
\lambda=0.1\left(\frac{1eV}{m_a}\right)\;(cm)
\end{equation}
If $m_a<1eV$ the length $\lambda \geq 0.1 cm$ and for samples of sizes
$\sim 1mm$ one can treat the field ${\bf B}_{eff}$ as a homogenious
one:
\begin{equation}\label{Field-hom}
{\bf B}_{eff}={\bf b}\sin
(m_at+\theta),\;\;
{\bf b}=2\kappa\sqrt{\rho_a}{\bf v}.
\end{equation}
Such an exotic quasimagnetic field with the amplitude about
$10^{-16} Gs$ can be picked up already in the present state of the art.
However, methods of its detection depend essentially on the axion mass
$m_a$ value being the frequency of the field's
${\bf B}_{eff}$ oscillations.

\section{Optical range detectors}
\subsection{Detection an axion-induced fluorescence}
Let us consider an atom or an ion with $L-S$ coupling and with more
than half full electronic shell. The field ${\bf B}_{eff}$
causes in this atom a weak mixing of different levels of the
fine structure. The amplitudes of this mixing oscillate in time
with the frequency $m_a$. Optical transitions between levels
of the fine structure will lead to an enhancement of optical
noise at the frequency $m_a$. It has a little value by itself.
However, it is of basic importance that the number of additional
optical photons depends on an orientation of the atom's angular
moment ${\bf J}$ with respect to the cosmological velocity
${\bf v}$.

It is convenient to describe the electron--condensat interaction
by the quantity $\Omega_a=4v\mu_a\sqrt{\rho_a}$ having the
frequency dimension. It is just the precession frequency
of the electron spin in the magnetic field ${\bf b}$. Let the mass
$m_a$ closely matches the energy gap $\Delta E$ between the
ground state $|JJ>\equiv |1>$ and the first exited level
$|J-1,J-1>\equiv |2>$ of the fine structure.
Then the amplitude of the optical noise can be estimated as:
\begin{equation}\label{Noise}
\dot{N}_\gamma\approx
\frac{\Omega_a^2\Gamma}{(\Delta E-m_a)^2+\Gamma^2}
|({\bf n}{\bf S})_{12}|^2.
\end{equation}
Here $\dot{N}_\gamma$ is the number of photons emitted per unit time,
$\Gamma$ is the width of the stirred level, ${\bf S}$ is the total
spin operator of the shell and the unit vector ${\bf n}$ is directed
along the cosmological velocity ${\bf v}$. Introducing the vector
${\bf J}$ equal to the expectation value of the total angular moment
we rewrite (\ref{Noise}) as:
\begin{equation}\label{Noise-J}
\dot{N}_\gamma\approx
\frac{\Omega_a^2\Gamma}{(\Delta E-m_a)^2+\Gamma^2}\,
\frac{SL}{2J^3}
({\bf n}\times{\bf J})^2,
\end{equation}
where $S$ and $L$ are the values of spin and orbital moment of the
electronic shell. In the resonance case $\dot{N}_\gamma$ is equal to:
\begin{equation}\label{Noise-re}
\dot{N}_\gamma\approx
\frac{\Omega_a^2}{\Gamma}
\frac{SL}{2J^3}
({\bf n}\times{\bf J})^2.
\end{equation}
We see that the noise amplitude depends on the angle between
${\bf J}$ and the "absolute" velocity ${\bf v}$.
The rotation of a detector containing such atoms or ions about
the axis perpendicular to ${\bf v}$ allows us to use the modulation
procedure. The latter improves essentially the sensitivity of the
experiment and allows one to detect the variations of the photonic
noise down to $10^{-4}Hz$. To estimate a possible sensitivity let us
consider the ytterbium ion $\hbox{Yb}^{+3}$ indroduced as a dopant
into the crystal $\hbox{CaF}_{2}$
or $\hbox{Y}_{3}\hbox{Al}_5\hbox{O}_{12}$. The suitable for our aims
optical transition $^2\hbox{F}_{5/2}$--$^2\hbox{F}_{7/2}$ has the
wavelength $\lambda=1.03\mu m$ and the width $\Gamma =10^4Hz$ at
the temperature 77K.
The modulation amplitude
$\Delta\dot{N}_\gamma$ of the noise intensity can be estimated as
\begin{equation}\label{deltaNoise-re}
\Delta\dot{N}_\gamma\sim
\frac{\Omega_a^2}{\Gamma}
n_i {\cal V},
\end{equation}
where $n_i$ is ion's concentration and $ {\cal V}$ is volume of the
crystal. If $n_i\approx 10^{22} cm^{-3}$ and $ {\cal V} =1cm^3$
the limitation $10^{-4}Hz$ on the measurable value of
$\Delta\dot{N}_\gamma$ gives a bottom limit of detectable quasimagnetic
field $b$ as $10^{-18} Gs$ or in terms of the frequency $\Omega_a$:
\begin{equation}\label{Om-lim}
\Omega_a\geq 10^{-11} Hz
\end{equation}
To re-express the limit (\ref{Om-lim}) in terms of the coupling
constant $\kappa$ one need to relate the local condensat density $\rho_a$
to the critical one $\rho_c$. Models taking into account the gravitation
effects \cite{Tu} give for $\rho_a$ in halo $\rho_a\approx 10^4 \rho_c$.
It leads to the limitation $\kappa<10^{-13}$ whereas the Sun physics
gives $\kappa<10^{-10}$.

Placing our sensitive sample in an external magnetic field we can fit the
axion mass using the Zeeman shift of the fine structure levels.

\subsection{Photo-induced decay of an axion condensate}
The decay of axion into two photons, like the decay of a neutral pion,
occurs through a triangle diagram with a virtual charged fermion
in a loop.

The lifetime of the axion whith respect to decay into two
photons is given by:
\begin{equation}\label{Lif-tim}
\tau_a = (m_{\tau}/m_a)^5~\tau_{\pi}
\end{equation}
where $m_a$  and $m_{\pi}$ are the masses of the axion and the pion,
and $\tau_a$ and $\tau_{\pi}$ are the corresponding lifetimes.
The lifetime of an axion with a mass of $1 eV$ is $3~10^{24}~s$.

Let us assume a coherent (parallel and monochromatic) stream of
axions with a "cosmological" velocity ${\bf v}$ respect to observer.
In the comoving frame of reference, an axion decays into two photons
with the same energy. They fly off in opposite directions and have
orthogonal polarization. Decay in which  one photon is emitted
along the direction in which the axion is moving, while the other
is emitted in opposite direction, lead to the greatest difference
in the photon energy as observed in the laboratory frame:
\begin{equation}\label{E-fot}
E_{max}=E_0(1+v/c),\;\; E_{min}=E_0(1-v/c)
\end{equation}
where $E_0 = m_ac^2/2$. Such decay are strongly suppressed by the
small solid angle, but these are the decays which are interest
here. We now assume that there is  an intense, coherent photon
beam which coincident with the axion beam. We assume that the
photon energy is $E_\gamma=E_{max}$ if the axion and photon
beams are headed in the same direction, while we have
$E_\gamma=E_{min}$ if they are headed in opposite directions.
As a result, the axion lifetime decreases in proportion to the
number of photons in the (laser) photon beam:
\begin{equation}\label{tau}
\tau=\tau_a/n_\gamma
\end{equation}
Here $n_\gamma$ is the number of photons in the laser beam.
The reason for this result is that a very significant
Bose-amplification factor arises in the probability for the decay of
an axion. This effect could be called a "photoinduced axion decay"
\cite{Liad}.
It is easy to see that the axion decay probability is given by
\begin{equation}\label{dprob}
p=(\tau_i/\tau_a)\tau_i\dot{n}_\gamma
\end{equation}
in the case of monochromatic parallel axion and photon beams.
Here $\tau_i=L/c$ is the duration of the interaction between
axion and photon beams, and $n_\gamma$ is the photon flux density,
and $\tau_a$ is the natural axion lifetime. With $\dot{n}_a$ and
$\dot{n}_\gamma$
as the flux densities of axions and photons, respectively, the flux
density of photons from induced decays is
\begin{equation}\label{photfl}
\dot{N}_\gamma=(\tau_i/\tau_a)\tau_i\dot{n}_\gamma\dot{n}_a
\end{equation}
One of photon from the axion decay has a frequency, a polarization,
and a direction which are the same as those in the laser beam.
The second photon propagates in the opposite direction and has a
polarazation orthogonal lo that of the laser beam. Its frequency is
different by twice the Doppler shift. This circumstance raises
the hope that an effective detector of axions with masses on the
order of $1 - 5 eV$ can be developed.

Let us assume that the axion mass is $2.5eV$, the laser power is
$10^4 w$, the interaction lenght is $L=10m$, and the axion flux
density is $10^{15}/cm^2/s$ (it correspond to the local axion
galo density).
We find the flux density of photon from induced axion decay to be
$\dot{N}_\gamma=10/cm^2/s$.

\section{Detectors for radiofrequencies range}
If the quasimagnetic field has frequency below $10^6$ Hz (what correspons
to $0<m_a<10^{-8}eV$) it is natural to use a detector with a ferromagnetic
rod as a sensitive body. Its magnetization can be read off by SQUID.
Detectors of such a kind have been used already in search of arion and
T-odd long-range forces \cite{Petja2}-\cite{Petja3}.
A probe consisting of high
quality paramagnet or antiferromagnet is placed inside a lead
superconducting screen of a layered structure. This screen suppresses
external magnetic fields down to $10^{-15}$Gs. The probe magnetization
is measured by SQUID magnitometer with a sensitivity
$10^{-6}\Phi_0/\sqrt{Hz}$ ($\Phi_0$ is the magnetic flux quantum).
Rotation of the detector enables us to use the modulation procedure
suppressing the noise. The reached in \cite{Petja2}-\cite{Petja3}
sensitivity
was $\sim 10^{-12}$Gs, but modifications allow to detect fields
about $\sim 10^{-15}$Gs.

In the range $10^8 Hz<m_a<10^{10}Hz$ ($m_a<10^{-4}eV$) one can
use the resonance axion-magnon conversion in magnetic ordered media
\cite{Petja4}- \cite{Kakh}.
Let us consider a resonator with a working mode
$TE_{110}$ and with a small spherical ferro- or antiferromagnetic
sample placed in its center. An external magnetic field is directed
in such a way that the averaged magnetization of the sample is
perpendicular to the magnetic component of the resonator's
eigenmode. The magnetic resonance frequency is fitted to be equal
to the eigenfrequency of the resonator. It provides a strong
coupling between the magnetic moment oscillations and the
electromagnetic ones. If $m_a$ coincides with this frequency
the spin waves will be exited resonancely by the axionic wind.
Electromagnetic oscillations coupled with such the waves can be
detected by a sensitive receiver.

Detailed discussion of the axion-magnon conversion and the corresponding
computations can be found in papers \cite{Petja5}, \cite{Kakh}. Here we present
the result only. If $P$ is a limiting value of intensity which our receiver
can detect, $M$ is the sample magnetization, $Q_f$ is the quality of the
ferromagnetic resonance and $H_0$ is the external magnetic field, then
the smallest detectable quasimagnetic field is equal to:
\begin{equation}\label{Lim}
b\approx \left(\frac{P}{m_a^4Q_f}\right)^{1/2}\frac{H_0}{M}\frac{1}{kL}
\end{equation}
and for $P=10^{-15}erg/c$, $m_a=10^{10}Hz$, $H_0\sim M \approx 10^3 Gs$
and $(kL)^2=10^3$
we obtain $b\approx 10^{-15}Gs$. The use of the antiferromagnet with a
large Dzyaloshinsky field (e.g. $Fe B O_3$) as a sensitive body can give
an additional factor $\sim 10^{-3}$ in the right hand side of (\ref{Lim})
(see \cite{Kakh}).

The idea to use ferromagnetic precession for detecting halo axions
has been proposed also in \cite{Barb}.

\section{Status of the axion-search program in BINP (Novosibirsk)}
We shortly describe principal features of BINP axion experiments
and their present status:

1. "Helioscope" - this detector is able to detect the solar axions
within the mass range $0 - 0.1 eV$. The experiment is based on
the effect of resonant axion-photon conversion in a strong
transverse magnetic field. The base of detector is a dipole
SC magnet (B=60kG, lenght-6m) for the coherent axion-photon conversion.
Its mass, with a criostat (LHe), is 10 tons. It is mounted on a
rotatable platform and is able to follow the Sun with accuracy
about 1arcmin. A proportional chamber registers the photons born
due to conversion, with energies $3 - 10 KeV$. The detector provides
sensitivity on axion-photon coupling constant less
then $10^{-10}GeV^{-1}$.

Status for 01.12.1994:

 --- The detector is mounted, its control and data acquisition system
   are ready.

 --- The vacuum and cryogenic test, at temperature of LN, were performed.

 --- The cryotronic power supply for SC-magnet (I=6KA) - not ready.

We hope to start data acquisition on summer 1995, and finish
the experiment for the end 1995.

2. "Helioscope-2" - search for solar axions of mass $1 - 5 eV$.
This detector is based on idea of light-induced axion decay in
the presence of intensive coherent beam of photons \cite{Liad}.
Design of detector: an UHF-power is storing in a superconducttir
waveguide with ends closed by mirror, reflecting UHF power and
transparent for X-rays. Semiconductor counter, for X-ray registration,
are interposed on the ends of waveguide.
The waveguide is placed into a liquid helium.
All it follows the Sun.

Status for 01.12.1994:

 --- The desugning of SC RF cavity and its exciting system are at
the stage of performing.

 --- The cryostat for SC cavity is ready.

 --- We are elaborating an alternative variant, working at room
 temperature.

3. "Haloscope-1" - search for galactic axions of masses about
 $10^{-5} - 10^{-4}eV$. Teoretically, such axions should provide
cosmological cold dark matter in form of the axion condensate.
The experiment aims to search for axionic dark matter using
quasimagnetic interaction of the axion with the electron spin.
It is based on the effect of resonant axion-magnon conversion
in magnetic ordered media \cite{Petja4}- \cite{Kakh}, \cite{Barb}.

Status for 01.12.1994:

  --- The data acquisition.

4. "Haloscope-2" - search on galactic axions.
The detector operating in axion mass range near 1 eV, which
is of interast in the context of the archion model  --- in this
mass range the archions should constitute the cold (or warm)
component of the cosmological dark matter \cite{Khl1}, \cite{Berez}.
The experiment  is based on the
resonant excitation effect of optical transitions between
the fine structure levels \cite{VKK}.  AYG monocrystall
is used, activated by $Yb^{+3}$ and placed in a magnetic field
(B = 70 kG) as a sensitive element.

Status of the "Haloscope-2" detector on 01.12.1994 --- we alredy have
for this project:

  --- Cryostat,

  --- SC magnet (70 kG),

  --- AIG samples.

5.  We are designed and investigation an parametric
axion-photon RF - transductor.

\vspace{2mm}

We are grateful to A.A.Anselm,  I.B.Khriplovich and  A.S.Yelkhovsky\\
for useful conversations and notices.
It is a please for us to acknowledge the support from the
COSMION foundation [P.V] and the Soros foundation [I.K.].

\end{document}